\documentclass[12pt]{iopart}
\usepackage{graphicx}

\begin{document}

\title{$^{77}$Se and $^{63}$Cu NMR studies of the electronic correlations in Cu$_x$TiSe$_2$ ($x=0.05, 0.07$)}

\author{L L Lumata$^{1}$, K Y Choi$^{2}$, J S Brooks$^{1}$, A P Reyes$^{1}$, P L Kuhns$^{1}$, G Wu$^{3}$ and X H Chen$^{3}$}

\address{$^{1}$Department of Physics  and National High Magnetic
Field Laboratory, Florida State University, Tallahassee, Florida
32310, USA}

\address{$^{2}$Department of Physics, Chung-Ang University, 221
Huksuk-Dong, Dongjak-Gu, Seoul 156-756, Republic of Korea}

\address{$^{3}$Hefei National Laboratory for Physical Science at
Microscale and Department of Physics, University of Science and
Technology of China, Hefei, Anhui 230026, People's Republic of
China}

 \ead{lloyd.lumata@utsouthwestern.edu}
\begin{abstract}
We report $^{77}$Se and $^{63}$Cu nuclear magnetic resonance (NMR)
investigation on the charge-density-wave (CDW) superconductor
Cu$_x$TiSe$_2$ ($x=0.05$ and 0.07). At high magnetic fields where
superconductivity is suppressed, the temperature dependence of
$^{77}$Se and $^{63}$Cu spin-lattice relaxation rates $1/T_{1}$
follow a linear relation. The slope of $^{77}1/T_{1}$ vs \emph{T}
increases with the Cu doping. This can be described by a modified
Korringa relation which suggests the significance of electronic
correlations and the Se 4\emph{p}- and Ti 3\emph{d}-band
contribution to the density of states at the Fermi level in the
studied compounds.
\end{abstract}


\section{Introduction}
One of the central issues in condensed matter physics is to
understand the interplay between diverse competing phases: for
example, antiferromagnetism and superconductivity in the
high-$T_c$ cuprates~\cite{Imada} and charge-density-wave (CDW)
order and superconductivity in transition metal
dichalcogenides~\cite{wilson}. Very recently, the CDW compound
TiSe$_2$ has been successfully doped by a Cu intercalation between
adjacent layers~\cite{morosan}. This allows for tuning the
electronic properties of Cu$_{x}$TiSe$_{2}$ by controlling the Cu
concentration $x$. The Cu doping drives the system to a
superconducting state around $x=0.04$. In this regard, the
Cu$_{x}$TiSe$_{2}$ system provides a unique opportunity to study
the competition between the CDW and superconducting phases in a
controlled way.

The Cu$_{x}$TiSe$_{2}$ compound has a layered structure with
trigonal symmetry where the $Ti$ atoms are in octahedral
coordination with Se. The Cu atoms occupy the positions between
the TiSe$_{2}$ layers which leads to an expansion of the lattice
parameters~\cite{morosan}. The parent compound TiSe$_2$ is one of
the first CDW compounds, where a commensurate CDW at
$T_{cdw}\sim200$ K with a wave vector $Q = (2a, 2a, 2c)$ was
reported~\cite{wilson,salvo}. The CDW transition is not driven by
the nesting of a Fermi surface. Instead it is either due to a
transition from a small indirect gap into a larger gap state with
a slightly different location in the Brillouin zone or due to a
particle-hole symmetry~\cite{morosan, Qian}. The CDW transition in
TiSe$_2$ can be suppressed continuously by a Cu doping while
superconductivity emerges with a maximum $T_{c} = 4.15$ K at
$x=0.08$.  The electronic structure has  been investigated by band
structure calculations and ARPES
measurements~\cite{Qian,Jeong,Jishi,Zhao}. The remarkable feature
is that the CDW order parameter competes microscopically with
superconductivity in the same band and that the parent compound
TiSe$_{2}$ is a correlated semiconductor and the Cu doping
enhances the density of states and raises the chemical potential.

Various measurements were also performed to further investigate
the role of the Cu doping~\cite{Syli,Li,GWu,Barath,Budko}. Thermal
conductivity suggests that this system exhibits conventional
\emph{s}-wave superconductivity due to the absence of a residual
linear term $\frac{\kappa_{0}}{T}$ at very low
temperatures~\cite{Syli}. The weak magnetic field dependence of
thermal conductivity indicates a single gap that is uniform across
the Fermi surface. Furthermore, it is suggested that the $4p$ Se
band is below the Fermi level and superconductivity is induced
because of the Cu doping into the $3d$ Ti bands~\cite{Syli}. An
optical spectroscopy investigation on Cu$_{0.07}$TiSe$_{2}$
reveals that the compound has a low carrier density and has an
anomalous metallic state because of the substantial shift of the
screened plasma frequency~\cite{Li}. This is corroborated by the
temperature-independent Hall coefficient $R_{H}$ observed in
heavily Cu-doped samples~\cite{GWu}. Raman scattering measurements
as a function of doping suggest that the $x$-dependent mode
softening is associated with the reduction of electron-phonon
couplings and the presence of a quantum critical point within the
superconducting region~\cite{Barath}.

In this paper we address the issue of electronic correlations in
this system as a function of Cu doping using $^{77}$Se and
$^{63}$Cu NMR. We studied two doping levels of Cu$_{x}$TiSe$_{2}$
which exhibit CDW and/or superconductivity at zero field: $x=0.05$
($T_{cdw}=90$ K, $T_{c}=1.6$ K) and $x=0.07$ ($T_{c}=3.5 $ K). At
high magnetic fields, superconductivity is suppressed and we find
that the spin-lattice relaxation rate is described by a modified
Korringa relation  with a doping-dependent Korringa factor. The
large values of the extracted interaction parameters signify the
importance of electron-electron interactions and/or the Se
\emph{p}- and Ti \emph{d}-band contribution for these materials.
This is a key ingredient to the superconductivity for these
compounds.

\begin{figure}[tbp]
\centering \linespread{1}
\par
\includegraphics[width=3.0 in]{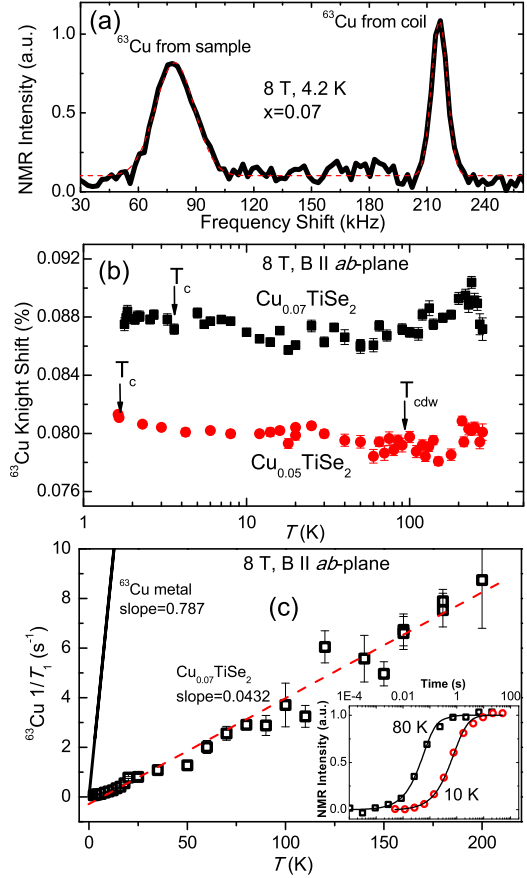}
\par
\caption{(a) Representative $^{63}$Cu NMR spectrum showing the two
peaks from the sample and the copper coil. (b) Temperature
dependence of $^{63}$Cu NMR Knight shift at 8 T for x=0.05 and
0.07. The arrows indicate the transition temperatures at zero
field. (c) $^{63}$Cu spin-lattice relaxation rate $1/T_{1}$ versus
temperature in Cu$_{0.07}$TiSe$_2$ at 8 T for
\textbf{B}$\parallel$\emph{ab}-plane. The solid line is the
temperature dependence of $1/T_1$ for $^{63}$Cu metal. Inset:
Representative longitudinal magnetization recovery curves at 10 K
and 80 K fitted by the master equation described in the text.}
\label{fig:1}
\end{figure}

\section{Experimental Details}
Single crystals of Cu$_{x}$TiSe$_{2}$ were grown by chemical
iodine-vapor transport method and characterized by x-ray,
transport, and susceptibility measurements~\cite{GWu}. The
plate-like samples with average thickness of $30~\mu$m along the
$c$-axis were cut in rectangular shapes with approximate
dimensions $\mathrm{4~mm \times 2~mm}$. Parallel stacks of
rectangular samples were placed in a rectangular-shaped copper
coil for a better filling factor with the magnetic field B applied
along the \emph{ab} plane (parallel with the plates) for NMR
measurements. Two doping levels ($x = 0.05, 0.07$) of
Cu$_{x}$TiSe$_{2}$ were studied under the same conditions.
$^{77}$Se and $^{63}$Cu NMR spectra, Knight shift, and
spin-lattice relaxation rates were obtained using a homebuilt NMR
spectrometer equipped with a high homogeneity 15 T sweepable
magnet. $T_1$ was measured by the recovery of the spin echo
intensity as a function of time after saturation.

\section{Results and Discussion}

\begin{figure}[tbp]
\centering \linespread{1}
\par
\includegraphics[width=3 in]{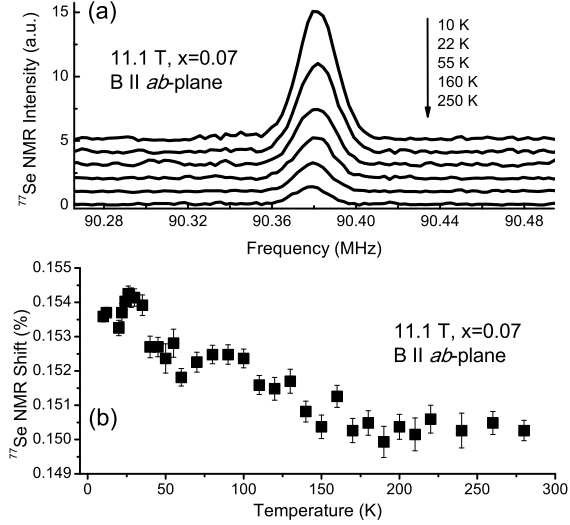}
\par
\caption{(a) Temperature dependence of $^{77}$Se NMR spectra of
Cu$_x$TiSe$_2$ ($x=0.07$) at 11.1~T for
\textbf{B}$\parallel$\emph{ab}-plane. (b) The corresponding
$^{77}$Se NMR shift as a function of temperature.} \label{fig:2}
\end{figure}

Figure~\ref{fig:1}(a) shows the representative  $^{63}$Cu
(spin-3/2, $\gamma_{n}=11.285$ MHz/T, 69.1 \% natural abundance)
NMR spectrum. The $^{63}$Cu NMR spectrum consists of two peaks,
which are assigned to the signals from the sample and copper coil
as indicated in the figure. The symmetric, sharp spectral shape of
the NMR signal indicates that the Cu site is not affected by CDW
modulation. The insensitivity of $^{63}$Cu nuclei to an electronic
state should be ascribed to the fact that the Cu atoms are located
in between the TiSe$_{2}$ layers which are bounded by van der
Waals interaction. Quadrupolar effects are expected for Cu nuclei
which occupies a site of non-cubic symmetry but the satellites are
not observable. The quadrupole frequency $\nu_{Q}$ for $^{63}$Cu
is usually on the order of 20-30 MHz \cite{Carter}, thus the
quadrupolar satellites are probably too broad and washed out due
to disorder since Cu is dilute in this system. As such, the
recovery of the longitudinal nuclear magnetization is
appropriately fitted by the master equation for a spin-3/2 central
transition using a single saturating pulse:
$M_{z}(t)=M_{0}[1-0.1e^{-(\frac{t}{T_{1}})}-0.9e^{-(6\frac{t}{T_{1}})}]$~\cite{Narath2}
(see Fig.\ref{fig:1}(c) inset). The temperature dependence of
$^{63}$Cu NMR Knight shift and spin-lattice relaxation rate
$1/T_{1}$ at 8~T is summarized in Figures~\ref{fig:1}(b) and (c).
The $^{63}$Cu Knight shift data exhibit very weak temperature
dependence with $K_{s}\approx 0.080$ \% for $x=0.05$ and
$K_{s}\approx 0.088$ \% for $x=0.07$, whereas in comparison
$^{63}$Cu metal has $K_{s}=0.2394$ \%~\cite{Carter}. There is no
hint that its temperature dependence scales to the magnetic
susceptibility. In addition, no anomalies are accompanied by the
CDW transition. This might be due to the fact that the external
field of 11.1 T fully suppresses the residual CDW state by
shifting the chemical potential and/or changing the electronic
density of states of Ti 3\emph{d} bands. The $^{63}$Cu
spin-lattice relaxation rate for $x=0.07$ varies linearly with
temperature, obeying Korringa behavior with a slope approximately
1/18 times that of a pure $^{63}$Cu metal.

Since the Se atoms are located on the layers, they probe the Fermi
surface and provide more information on electronic correlations in
this material. Figure~\ref{fig:2} displays temperature dependence
of $^{77}$Se (spin-1/2, $\gamma_{n}=8.13$ MHz/T, 7.5 \% natural
abundance) spectra and NMR shift of the $x=0.07$ compound at
B=11.1~T. The $^{77}$Se NMR signal shows the symmetric, sharp
spectral shape as well. The NMR shift is constant from 280 K to
150 K but increases monotonically upon further cooling. We note
that the room temperature NMR shift $K$ for the Cu$_x$TiSe$_2$
samples ($K\approx 0.141~\%$ for $x=5~\%$ and $K\approx 0.15~\%$
for $x=7~\%$) are close to the reported NMR shift values of the
selenide compounds TiSe$_2$ ($K_{iso}=0.111~\%$)~\cite{Dupree} and
TaSe$_2$ ($K_{iso}=0.162~\%$)~\cite{Borsa}.

In Figure~\ref{fig:3} we compare temperature dependence of
$1/T_{1}$ of Cu$_x$TiSe$_2$ ($x=0.05,0.07$) single crystals with
that of a $^{77}$Se metal~\cite{Carter} and powdered
TiSe$_{2}$~\cite{Dupree}. A single-exponential fitting equation
$M_{z}(t)=M_{0}[1-e^{-\frac{t}{T_{1}}}]$ was used to extract the
$T_{1}$ values from the magnetization recovery curves. It is
apparent that an increase of the Cu doping concentration leads to
an enhancement of the relaxation rate as well as an increase of
the slope of the Korringa-like behavior. The inset of
Fig.~\ref{fig:3} shows the non-linear $1/T_{1}$ vs T behavior of
the parent compound TiSe$_{2}$ and the subsequent linear behavior
upon Cu doping which suppresses the CDW state. In order to
understand the underlying mechanism of the observed $1/T_{1}$
behaviors we invoke a model involving a Korringa factor
$K(\alpha)$ due to the effect of electron-electron interaction in
metals and alloy compounds developed by Moriya~\cite{Moriya} and
later refined by Narath and Weaver~\cite{Narath}.

In simple metals, the dominant relaxation mechanism of nuclear
spins is due to their interaction with conduction electron spins.
This is given by the Korringa relation $1/T_{1} \propto T$ wherein
only Fermi contact interaction is taken into account for
noninteracting electrons. However electron-electron interactions
cannot be neglected especially for metal alloys. The Coulomb
interaction has an effect on the magnetic susceptibility which is
proportional to the Knight shift K$_{s}$~\cite{Moriya, Narath}.
The internal magnetic field experienced by a nucleus is given by
$H_{loc} = \langle H_{loc}\rangle +\delta H$ where $\langle
H_{loc}\rangle$ is the average internal magnetic field and $\delta
H$ is the fluctuating magnetic field. The nuclear spin lattice
relaxation rate is a measure of the perpendicular fluctuating
magnetic field $\delta H_{\perp}$ given by $1/T_{1} =
(\frac{\gamma^{2}}{2})\int_{-\infty}^{+\infty}dt \exp{(i\omega
t)}\langle \delta H_{\perp}(t)\delta H_{\perp}(0)\rangle $. In
metals with strictly non-interacting electron gas, the relaxation
rate is proportional to the temperature and square of the Knight
shift \cite{Moriya}, known as the Korringa relation:

\begin{figure}[tbp]
\centering \linespread{1}
\par
\includegraphics[width=3.5 in]{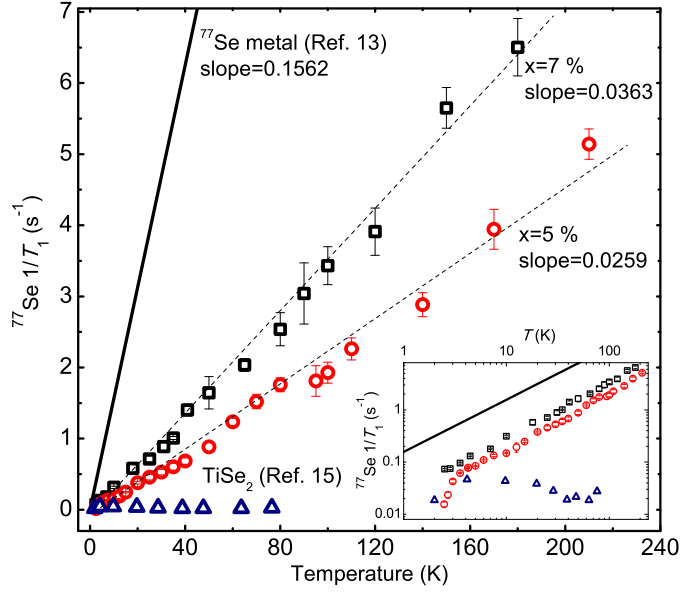}
\par
\caption{Comparison of temperature dependence of $^{77}$Se
spin-lattice relaxation rates $1/T_{1}$ for $^{77}$Se metal
(Ref.\cite{Carter}), Cu$_x$TiSe$_2$ (x=0.05, 0.07) single crystals
from this work, and the powdered parent compound TiSe$_2$ from
Ref.~\cite{Dupree}. Inset: log-log plot of $^{77}1/T_{1}$ vs
\emph{T}.}\label{fig:3}
\end{figure}

\begin{equation}
1/T_{1} = (\frac{4\pi
k_{B}T}{\hbar})\frac{\gamma_{n}^{2}}{\gamma_{e}^{2}}K_{s}^{2}
\label{eq:korr1}
\end{equation}

\noindent where $\gamma_n$, $\gamma_{e}$ are the nuclear and
electron gyromagnetic ratios, respectively. Moriya modified the
Korringa relation by incorporating the factor $K(\alpha)$ that
includes the effect of a $\delta$-function-potential
electron-electron interaction~\cite{Moriya}:

\begin{equation}
1/T_{1} = (\frac{4\pi
k_{B}T}{\hbar})\frac{\gamma_{n}^{2}}{\gamma_{e}^{2}}K_{s}^{2}K(\alpha)
\label{eq:korr}
\end{equation}

\noindent where the Korringa factor
$K(\alpha)=2\int_{0}^{1}\frac{(1-\alpha)^{2}xdx}{[1-\alpha
G(x)]^{2}}$ is a function of the interaction parameter $\alpha$, a
measure of electron-electron interaction.  Here $\alpha =
\frac{3\bar{v}n}{2\varepsilon_{F}}$ and $G(x)
=(\frac{1}{2})[1+\frac{1-x^{2}}{2x}\log{|\frac{1+x}{1-x}|}]$ from
the free electron approximation with $x = \frac{q}{2k_{F}}$ where
$\bar{v}$ is a constant, \emph{n} is the number of electrons,
\emph{q} is the wavevector, and $\varepsilon_{F}$, $k_{F}$ are the
Fermi energy and wave number, respectively \cite{Moriya}. The
value of $K(\alpha)$ ranges from 0 to 1. For $\alpha =0$,
$K(\alpha) = 1$ which gives us Eqn.(\ref{eq:korr1}). A corrected
version was later formulated by Narath appropriate for alkali and
noble metals where the function potential has a non-zero
interaction range less than the atomic radius~\cite{Narath}.

Before delving into the Korringa factor calculation, we note that
the orbital shift contribution cannot, in the present case, be
accurately separated from the spin part. Typically, the orbital
contribution can be unambiguously determined in the
superconducting state where the spin part of the shift vanishes at
low temperature due to pairing. However, access of NMR in the
superconducting state, believed to be of BCS type~\cite{Syli}, is
hampered by the low upper critical field ($H_{c2}\sim 1$
T)~\cite{morosan} of these compounds. We can however make some
estimates based on the NMR data from the parent compound
TiSe$_2$~\cite{Dupree}. The total NMR shift $K$ can be expressed
as:

\begin{equation}
K=K_{se}+K_{cp}+K_{vv}- \delta_{chem}
\end{equation}

\noindent where $K_{se}$ is the \emph{s}-electron contact term,
$K_{cp}$ is the \emph{p}-electron core polarization term, $K_{vv}$
is the Van Vleck orbital contribution, and $\delta_{chem}$ is the
chemical shift. Experimental estimates from a previous $^{77}$Se
NMR study on TiSe$_2$ \cite{Dupree} at room temperature yield
$K_{T}\approx 0.111\%$, $K_{se}\approx 0.03\%$, $|K_{cp}|\approx
0.01\%$ thus the temperature-independent part
$(K_{vv}-\delta_{chem})$ is 0.071\%. Since Cu doping into the
TiSe$_2$ layers increases the density of states (DOS) at the Fermi
level, this will increase $K_{se}$ and $K_{cp}$, but the terms
$K_{vv}$ and $\delta_{chem}$ should be unaffected since they do
not depend on DOS. We can then estimate the spin-part of the total
shift for Cu$_x$TiSe$_2$ system to be approximately ($K-0.071\%$).
This gives us the high temperature Knight shift values of 0.07\%
for $x=5\%$ ($K=0.141\%$) and 0.079\% for $x=7\%$ ($K=0.15\%$)
doping level (see Table~\ref{tab:kalpa}).

\begin{table}
\begin{center}
\caption{Korringa factor $K(\alpha)$ and interaction parameter
$\alpha$ due to electron-electron interactions based on $^{77}$Se
NMR data in Cu$_{x}$TiSe$_{2}$ ($x=0.05, 0.07$) and
TiSe$_{2}$~\cite{Dupree} at room temperature.}
\begin{tabular}{cccccc}
\hline\hline
Cu$_{x}$TiSe$_{2}$ & $1/T_{1}T$ $\mathrm{(s \cdot K)^{-1}}$ & $K_{s} (\%)$ & $K(\alpha)$ & $\alpha$ (Moriya) & $\alpha$ (Narath)\\
\hline
$x=0\%$~\cite{Dupree} & 0.0005 & 0.03 & 0.0401 & 0.955 & 0.99\\
$x=5\%$ & 0.0259 & 0.07 & 0.382 & 0.535 & 0.812\\
$x=7\%$ & 0.0363 & 0.079 & 0.42 & 0.505 & 0.782\\
\hline
\end{tabular}\label{tab:kalpa}
\end{center}
\end{table}

\begin{figure}[tbp]
\centering \linespread{1}
\par
\includegraphics[width=3.5 in]{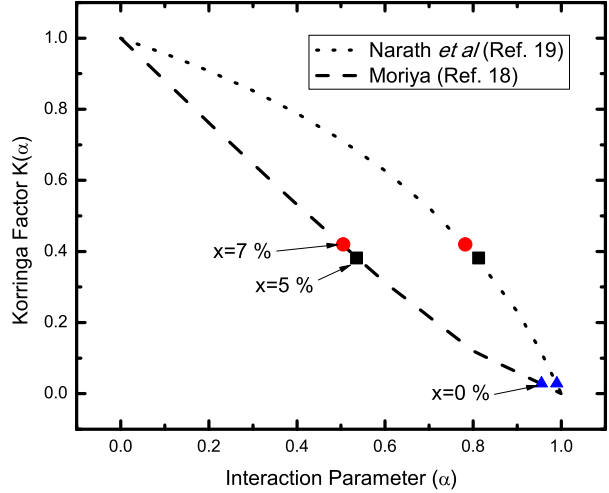}
\par
\caption{Korringa factor $K(\alpha)$ vs interaction parameter
$\alpha$ from models set by Moriya and a modified version from
Ref.~\cite{Narath}. The arrows indicate that the corresponding
interaction parameter $\alpha$ for Cu$_{x}$TiSe$_{2}$
($x=5~\%,~7~\%$) and TiSe$_{2}$ ($x=0~\%$) \cite{Dupree} based on
the K($\alpha$) values summarized in Table I.} \label{fig:4}
\end{figure}

The Korringa factors $K(\alpha)$ for Cu$_x$TiSe$_2$
($x=0\%\cite{Dupree},~5\%,~7\%$) system were calculated using
Eqn.(\ref{eq:korr}). By comparing our data to the enhanced
Korringa factor $K(\alpha)$ versus interaction parameter $\alpha$
calculated from Ref.\cite{Moriya} [see Fig.~\ref{fig:4}], we are
able to determine the interaction parameter $\alpha$. The
calculated values of $K(\alpha)$ and $\alpha$ are summarized in
Table~\ref{tab:kalpa}. First we revisit the $^{63}$Cu data in
Fig.\ref{fig:1} which yield $K(\alpha)=0.21$, $\alpha=0.71$ for
$x=7$ \% and $K(\alpha)=0.51$, $\alpha=0.42$ for a pure $^{63}$Cu
metal. The fact that $K(\alpha,^{63}Cu)>K(\alpha,x=7~\%)$
indicates that the exchange enhancement in the Cu metal is more
dominant than in the Cu$_{0.07}$TiSe$_2$ compound. For the
$^{77}$Se data, $K(\alpha)\approx0.38$, $\alpha=0.54$ for $x=5$ \%
and $K(\alpha)\approx0.42$, $\alpha=0.51$ for $x=7$ \% doping. The
large value of $\alpha$ highlights the significance of
electron-electron interactions. In comparison, data from the
parent compound TiSe$_2$ \cite{Dupree} at room temperature (see
Table~\ref{tab:kalpa}) yield $K(\alpha)\approx0.040$,
$\alpha=0.96$. The small value of $K(\alpha)$ for TiSe$_2$ may
suggest strong ferromagnetic correlations \cite{Shastry} but this
compound is viewed as a correlated semiconductor which may warrant
its deviation from this theoretical interpretation \cite{Hanoh}.
For Cu$_{x}$TiSe$_{2}$ system, the result
$K(\alpha,x=7\%)>K(\alpha,x=5\%)>K(\alpha,x=0\%)$ indicates that
Cu intercalation monotonically increases the exchange-enhancement
in the Se site. In addition to correlation effects, the
contribution of the density of states (DOS) should also be
considered in the observed relaxation behavior. Here
Cu$_x$TiSe$_2$ compounds evolve into better metals as evidenced by
decreasing resistivity and increasing susceptibility as Cu doping
is increased \cite{morosan, GWu}. This points to the fact that
increasing Cu content introduces carriers into the TiSe$_2$
conduction band, increasing the DOS at the Fermi level
\cite{morosan} thereby enhancing the relaxation rates in agreement
with our experimental data.

\section{Conclusion}
We have presented $^{77}$Se and $^{63}$Cu NMR studies of
Cu$_x$TiSe$_2$ ($x=5\%,~7\%$). We find that $^{63}$Cu Knight shift
shows weak variations with temperature and the spin-lattice
relaxation rate $^{77}1/T_{1}$ follows a modified Korringa
relation. The observed Korringa factor was discussed in terms of
Moriya's exchange-enhanced relaxation model and density of states
contribution. Our results suggest that enhanced electronic
correlations and Se \emph{p}- and Ti \emph{d}-band play a key role
in inducing superconductivity in the Cu$_x$TiSe$_2$ compounds.

\section{Acknowledgement}
This work is supported by the US National Science Foundation (NSF)
DMR grant 0602859, Priority Research Center Program (2009-0093817)
and Basic Science Research Program (2009-0077376) of Korea NRF,
and the National Basic Research Program for China 2006CB922005.
The experiments were performed at the National High Magnetic Field
Laboratory which is supported by NSF DMR-0084183, by the State of
Florida, and by the Department of Energy.


\section*{References}
\begin{thebibliography}{100}

\bibitem{Imada}
Imada M, Fujimori A and Tokura Y 1998 {\it Rev. Mod. Phys.}
\textbf{70} 1039


\bibitem{wilson}
Wilson J A, Di Salvo F J and Mahajan S 1975, {\it Adv. Phys.}
\textbf{24} 117

\bibitem{morosan} Morosan E, Zandbergen H W, Dennis B S, Bos G W G, Onose Y, Klimczuk T,
Ramirez A P, Ong N P and Cava R J {\it Nature Phys.} \textbf{2}
544

\bibitem{salvo}
Di Salvo F J, Moncton D E and Waszczak J V 1976 {\it Phys. Rev.} B
\textbf{14} 4321

\bibitem{Qian} Qian D, Hsieh D, Wray L, Xia Y, Wang N L, Morosan E, Cava R J and Hasan M Z 2007 {\it Phys. Rev. Lett.} \textbf{98} 117007

\bibitem{Jeong}
Jeong T and Jarlborg T 2007 {\it Phys. Rev.} B \textbf{76} 153103

\bibitem{Jishi}
Jishi R A and Alyahyaei H M 2008 {\it Phys. Rev.} B \textbf{78}
144516



\bibitem{Zhao} Zhao J F, Ou H W, Wu G, Xie B P, Zhang Y, Shen D W, Wei J, Yang L X, Dong J K, Arita M,
Namatame H, Taniguchi M, Chen X H and Feng D L 2007 {\it Phys.
Rev. Lett.} \textbf{99} 146401



\bibitem{Syli} Li S Y, Wu G, Chen X H and Taillefer L 2007 {\it Phys. Rev. Lett.} \textbf{99} 107001


\bibitem{Li} Li G, Hu W Z, Qian D, Hsieh D, Hasan M Z, Morosan E, Cava R J and
Wang N L 2007 {\it Phys. Rev. Lett.} \textbf{99} 027404



\bibitem{GWu} Wu G, Yang H X, Zhao L, Luo X G, Wu T, Wang G Y and
Chen X H 2007 {\it Phys. Rev. B} \textbf{76} 024513



\bibitem{Barath} Barath H, Kim M, Karpus J F, Cooper S L, Abbamonte P, Fradkin E, Morosan E and
Cava R J 2008 {\it Phys. Rev. Lett.} \textbf{100} 106402

\bibitem{Carter} Carter G C, Bennett L H and Kahan D J (eds) 1977 \emph{Metallic Shifts in NMR} (Oxford: Pergamon)

\bibitem{Budko} Budko S L, Canfield P C, Morosan E, Cava R J and Schmiedeshoff G M 2007 {\it J. Phys.: Condens. Matter} \textbf{19} 176230

\bibitem{Narath2} Narath A 1967 {\it Phys. Rev.} \textbf{162} 320

\bibitem{Dupree}
Dupree R, Warren W W Jr and DiSalvo F J 1977 {\it Phys. Rev.} B
\textbf{16} 1001

\bibitem{Borsa} Borsa F and Torgeson D R 1976 \textit{Bull. Am. Phys.
Soc.} \textbf{21} 262

\bibitem{Moriya}
Moriya T 1963 {\it J. Phys. Soc. Japan} \textbf{18} 516

\bibitem{Narath}
Narath A and Weaver H T 1968 {\it Phys. Rev.} \textbf{175} 373

\bibitem{Shastry} Shastry B S and Abrahams E 1994 {\it Phys. Rev.
Lett.} \textbf{72} 1933

\bibitem{Hanoh} Lee H O and Choi H Y 2000 {\it Phys. Rev.} B
\textbf{62} 15120





\end{thebibliography}
\end{document}